\documentclass[11pt,twoside]{article}
\usepackage{asp2010}
\usepackage{amssymb}
\usepackage{graphicx,mathptm} 
\usepackage{subfigure}
\newcommand{\be}{\begin{equation}}
\newcommand{\ee}{\end{equation}}
\newcommand{\bea}{\begin{eqnarray}}
\newcommand{\eea}{\end{eqnarray}}

\resetcounters

\begin{document}

\title{Nonlinear transport of Cosmic Rays in turbulent magnetic field}
\author{Huirong Yan and Siyao Xu
\affil{$^1$KIAA, Peking University, Beijing 100871, China}}

\begin{abstract}
Recent advances in both the MHD turbulence theory and cosmic ray observations call for revisions in the paradigm of cosmic ray transport. We use the models of magnetohydrodynamic turbulence that were tested in numerical simulation, in which turbulence is injected at large scale and cascades to to small scales. We shall present the nonlinear results for cosmic ray transport, in particular, the cross field transport of CRs and demonstrate that the concept of cosmic ray subdiffusion in general does not apply and the perpendicular motion is well described by normal diffusion with $M_A^4$ dependence. Moreover, on scales less than injection scale of turbulence, CRs' transport becomes super-diffusive. Quantitative predictions for both the normal diffusion on large scale and super diffusion are confronted with recent numerical simulations. Implication for shock acceleration is briefly discussed. \end{abstract}

\section{Introduction}
The propagation and acceleration
of cosmic rays (CRs) is governed by their interactions
with magnetic fields. Astrophysical magnetic fields are turbulent and, 
therefore, the resonant and non-resonant (e.g. transient time damping, or TTD)
interaction of cosmic rays with MHD turbulence is the accepted
 principal mechanism to scatter and isotropize
cosmic rays \citep[see][]{Schlickeiser02}. In addition, efficient scattering is essential for the acceleration of cosmic rays. 
For instance, scattering of cosmic rays back into the shock is a
vital component of the first order Fermi acceleration \citep[see][]{Longairbook}. At the same time, stochastic acceleration by turbulence is 
entirely based on scattering. The dynamics of cosmic rays in MHD turbulence holds one of the keys to high energy astrophysics and related problems. 

At present, the propagation of the CRs is an advanced theory, which makes
use both of analytical studies including nonlinear formalism and numerical simulations. However,
these advances have been done within the turbulence paradigm which
is being changed by the current research in the field.
Instead of the empirical 2D+slab model of turbulence, numerical
simulations suggest anisotropic Alfv\'en and slow modes (an analog of 2D, but not an
exact one, as the anisotropy changes with the scale involved) + fast modes \citep{CL02_PRL}.

Propagation of CRs perpendicular to mean magnetic field
is another important problem for which one needs to take into account both large and small scale interactions in tested models of turbulence. Indeed, if one takes only the diffusion along the magnetic field line and field line random walk \citep[FLRW][]{Jokipii1966, Jokipii_Parker1969, Forman1974}, compound (or subdiffusion) would arise. Whether the subdiffusion is realistic in fact depends on the models of turbulence chosen \citep{YL08}. In this paper we again present our understandings to this problem within the domain of numerically tested models of MHD turbulence.

In what follows, we discuss the nonlinear cosmic ray transport in \S2. In \S3, we shall study the perpendicular transport of cosmic rays. We shall also discuss the issue of super-diffusion and its implication in \S4. Summary is provided in \S5.

\section{Cosmic ray scattering in nonlinear theory}

\subsection{Model of MHD turbulence}

Obviously the model of turbulence determines the CR transport properties. Recent years have seen substantial progress in the understanding of MHD turbulence, which is different from the conventionally adopted slab model and isotropic Kolmogorov model. It has been demonstrated by a number of studies that MHD turbulence can be decomposed into three modes: Alfven, slow and fast modes. Among them, Alfven and slow modes can be well described by the \cite[][henceforth GS95]{GS95} model and exhibit scale-dependent anisotropy with respect to the {\em local} magnetic field \citep{LV99, CL02_PRL}.  Fast modes, however, are isotropic \citep{CL02_PRL}.

The anisotropy leads to strong suppression of gyroresonance of CRs by the Alfvenic turbulence, which was the favorite modes in earlier literatures. 
The studies in \cite{YL02, YL04} demonstrated with quasilinear theory (QLT), nevertheless, that the scattering by fast modes dominates in most cases in spite of the damping\footnote{On the basis of weak turbulence theory, \cite{Chandran2005} has argued that high-frequency 
fast waves, which move mostly parallel to magnetic field, generate Alfven waves also moving mostly parallel to magnetic field. We expect
that the scattering by thus generated Alfven modes to be similar to the scattering by the fast modes created by them. Therefore
we expect that the simplified approach adopted in \cite{YL04} and the papers that followed to hold.}.

\subsection{Nonlinear theory}

The conclusion that fast modes dominates the gyroresonance interaction over Alfvenic turbulence is also confirmed in nonlinear theory (NLT, see YL08). Efficient gyroresonance itself does not ensure finite mean free path/diffusion coefficient for the CRs because particles cannot be scattered through gyroresonance at $90^\circ$. This is the longstanding $90^\circ$ problem, which is intrinsically associated with the QLT approach.  

Indeed, many attempts have been made to improve the QLT and various non-linear
 theories have been attempted (see \citealt{Dupree:1966}, V\"olk 1973, 1975, 
Jones, Kaiser \& Birmingham 1973, Goldstein 1976)  \citep[see][]{Matthaeus:2003},  \citep{Qin_NLT, LeRoux:2007}. An important step was taken in Yan \& Lazarian (2008), where non-linear effect was included in treating CR scattering in the type of MHD turbulence that are tested by numerical simulations. The results have been applied to both solar flares (Yan, Lazarian \& Petrosian 2008) and grain acceleration \citep{HLS12}. Below, we introduce the nonlinear theory and their applications to particle transport in turbulence based on the results from Yan \& Lazarian (2008).

The basic assumption of QLT is that the particles' orbit is unperturbed until a scattering event occurs. In reality, however, particle pitch angle  does change because of magnetic field compression according to the first adiabatic invariant $v_\bot^2/B$, where $B$ is the total strength of the magnetic field \citep[see][]{Landau:1975}. Since B varies in turbulent field, so are the projections of the particle speed $v_\bot$ and $v_\|$.
 This results in broadening of the resonance from the delta function $\pi\delta(k_{\parallel}v_{\parallel}-\omega\pm n\Omega)$ in QLT to 
\be
R_n(k_{\parallel}v_{\parallel}-\omega\pm n\Omega)\simeq\frac{\sqrt{\pi}}{|k_\||v_\bot \sqrt{M_A}}\exp\left[-\frac{(k_\|v \mu-\omega+n\Omega)^2}{k_\|^2v_\bot^2M_A}\right]
\label{resfunc}
\ee
where $M_A\equiv \delta V/v_A=\delta B/B_0$ is the Alfv\'enic Mach number and $v_A$ is the Alfv\'en speed. Note that Eqs.~(\ref{resfunc}) are generic, and applicable to both incompressible and compressible medium. 

For gyroresonance ($n=\pm 1,2,...$), the result is similar to that from QLT for $\mu\gg \Delta \mu=\Delta v_\|/v$. In this limit, Eq.(\ref{resfunc}) represents a sharp resonance and becomes equivalent to a $\delta$-function.  On the other hand, the dispersion of the $v_\parallel$ means that CRs with a much wider range of pitch angle can be scattered by the compressible modes through the TTD interaction 
($n=0$), which is marginally affected by the anisotropy and much more efficient than the gyroresonance. In QLT, the projected particle speed should be comparable to phase speed of the magnetic field compression according to the $\delta$ function for the TTD resonance. This means that only particles with a specific pitch angle 
can be scattered. With the resonance broadening, however, wider range of pitch angle can be scattered through TTD, including $90^\circ$. Finite mean free path is therefore ensured in NLT \citep{YL08}.


\subsection{Comparison with test particle simulations}

We run test particle simulations with turbulence generated from 3D MHD simulations. Test particle simulation has been used to study CR scattering and
transport \citep{Giacalone_Jok1999, Mace2000}. The aforementioned studies, however, used synthetic
data for turbulent fields, which have several disadvantages.
Creating synthetic turbulence data which has scale-dependent
anisotropy with respect to the local magnetic field (as observed
in \citealt{CV00} and \citealt{MG01}) is difficult
and has not been realized yet.  Also,
synthetic data normally use Gaussian statistics and
delta-correlated fields, which is hardly appropriate for
description of strong turbulence. 

\begin{figure*}[htbp]
   \centering
   \subfigure[]{
   \includegraphics[width=0.45\textwidth]{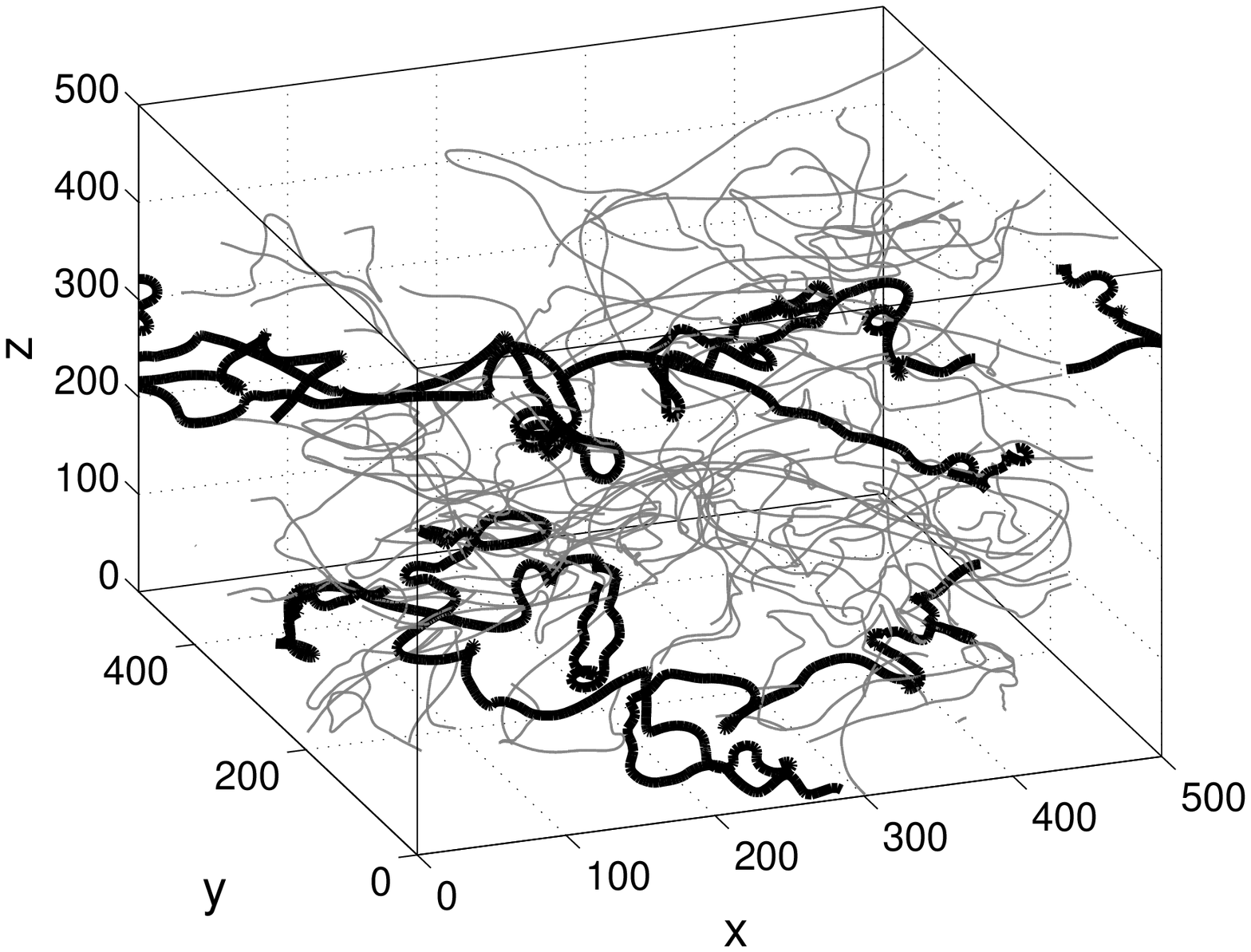}}
   \subfigure[]{
   \includegraphics[width=0.45\textwidth]{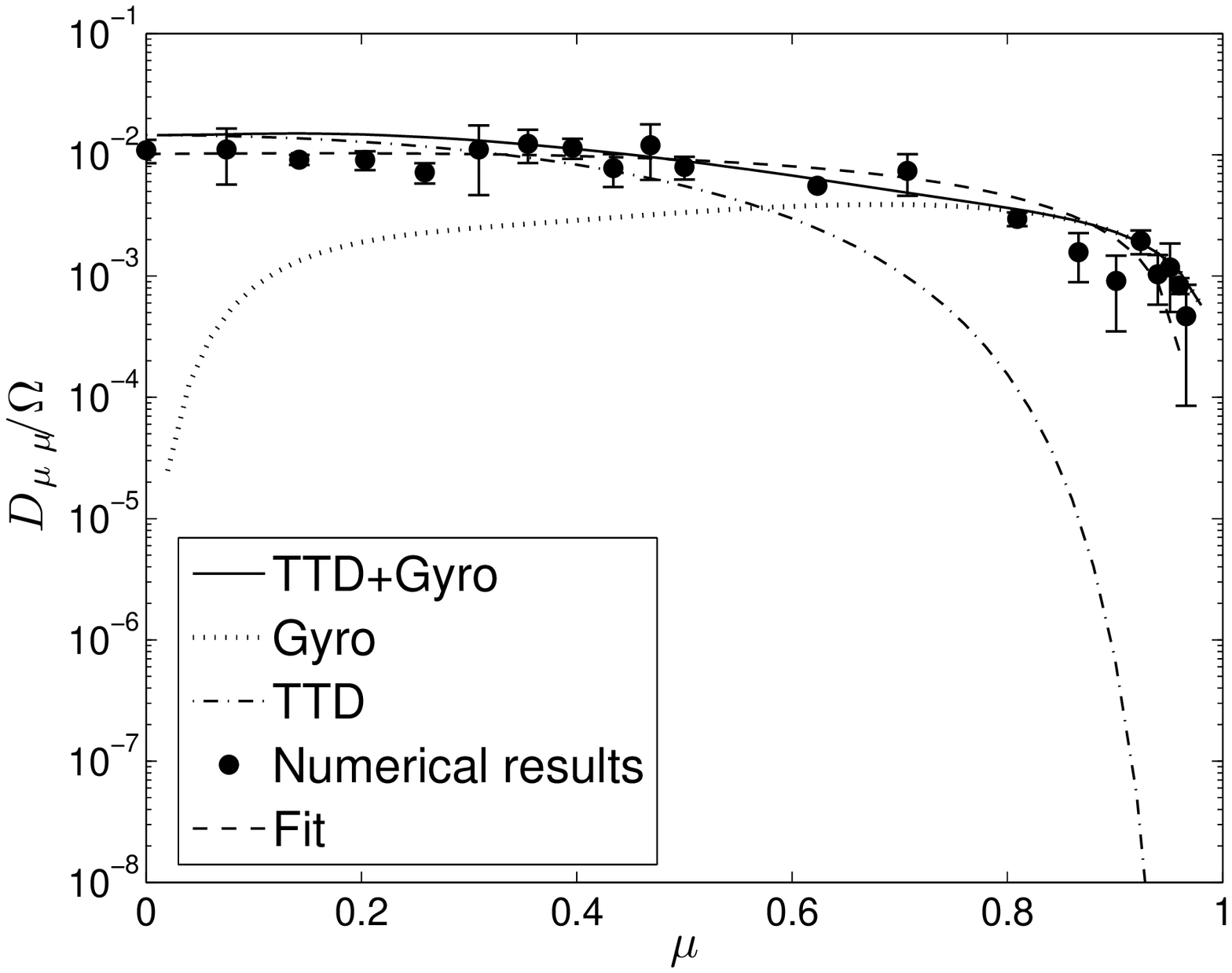}}
   \caption{\small (a) Particle trajectories (thick black lines) in MHD turbulence. The thin gray lines display the magnetic field stream lines; (b) dimensionless CR scattering coefficient $D_{\mu
    \mu}/\Omega$ vs the pitch angle $\mu$. It is dominated by TTD resonant mirror interaction with compressible modes\citep*[from][]{Xu_Yan}.}
\label{xx_yy}
\end{figure*}

Using the results of direct numerical MHD simulations as the input data, \cite{Xu_Yan} performed test particle simulations. The results show good correspondence with the analytical predictions. As shown in Fig.\ref{xx_yy}a, particles' trajectories are traced in the turbulent magnetic field.  The scattering coefficient shows the same pitch angle dependence as that predicted in \cite{YL08}, namely the scattering is most efficient for large pitch angles due to the TTD mirror interaction and dominated by gyroresonance at small pitch angles (see Fig.\ref{xx_yy}b).  Both interactions are necessary and complementary to each other in scattering CRs. 

\section{Perpendicular transport}

In this section we deal with the diffusion perpendicular
to {\it mean} magnetic field. Both observation of Galactic CRs and solar wind indicate that the diffusion of CRs perpendicular to magnetic field is comparable to parallel diffusion \citep[]{Giacalone_Jok1999, Maclennan2001}. 

The perpendicular transport is slow if particles are restricted to the magnetic field 
lines and the transport is solely due to the random walk of field 
line wandering \citep[see][]{Kota_Jok2000}. 
In the three-dimensional turbulence, field lines are diverging away due to shearing by Alfv\'en modes \citep[see][]{LV99, Lazarian06}.
 Since the Larmor radii of CRs are much larger than the minimum scale of eddies $l_{\bot, min}$, field lines within the CR Larmor orbit are effectively diverging away owing to shear by Alfv\'enic turbulence.
The cross-field transport  thus results from 
the deviations of field lines at small scales,
as well as field line random walk at large scale ($>{\rm min}[L/M^3_A,L]$), where $L, M_A$ are the injection scale and Alfv\'enic Mach number of the turbulence, respectively.

\subsection{Perpendicular diffusion on large scale}

{\it High $M_A$ turbulence}: High $M_A$ turbulence corresponds to the field that is easily bended by
hydrodynamic motions at the injection scale as the hydro energy at the
injection scale is much larger than the magnetic energy, i.e.
$\rho V_L^2\gg B^2$. In this case
magnetic field becomes dynamically important on a much smaller scale, i.e. the 
scale $l_A=L/M_A^3$. If the parallel mean free path of CRs $\lambda_\|\ll l_A$, the stiffness of B field is negligible so that the perpendicular diffusion coefficient is the same as the parallel one, i.e., $D_\bot=D_\|$. If $\lambda_\|\gg l_A$, the
 diffusion is controlled by the straightness of the field lines, and $
D_\bot=D_{\|}\approx 1/3l_Av,~~~M_A>1,~~~\lambda_{\|}>l_A.
\label{dbb}
$ The diffusion is isotropic if scales larger than $l_A$ are
concerned. In the opposite limit $\lambda_{\|}<l_A$, naturally, a result for
isotropic turbulence, namely, $D_{\bot}= D_{\|}\sim 1/3 \lambda_{\|} v$ holds. 

{\it Low $M_A$ turbulence}: For strong magnetic field, i.e. the field that cannot be easily bended at
the turbulence injection scale, individual magnetic field lines are aligned
with the mean magnetic field. The diffusion in this case is anisotropic.
If turbulence is injected at scale $L$ it stays 
weak for the scales larger than $LM_A^2$ and it is 
strong at smaller scales. Consider first the case of $\lambda_\|>L$.
The time of the individual step is $L/v_\|$, then $D_\perp\approx 1/3Lv M_A^4, ~~~M_A<1,~~~ \lambda_\|>L.$
This is similar to the case discussed in the FLRW model (Jokipii 1966). However, we obtain the dependence of $M_A^4$ instead of their $M_A^2$ scaling. In the opposite case of $\lambda_\|<L$, the perpendicular diffusion coefficient is $
D_{\bot}\approx D_{\|}M_A^4,
\label{diffx}$ which coincides with the result
obtained for the diffusion of thermal electrons in magnetized plasma \citep{Lazarian06}.

\begin{figure}[htbp]
     \centering
   \subfigure[]{
   \includegraphics[width=0.48\textwidth]{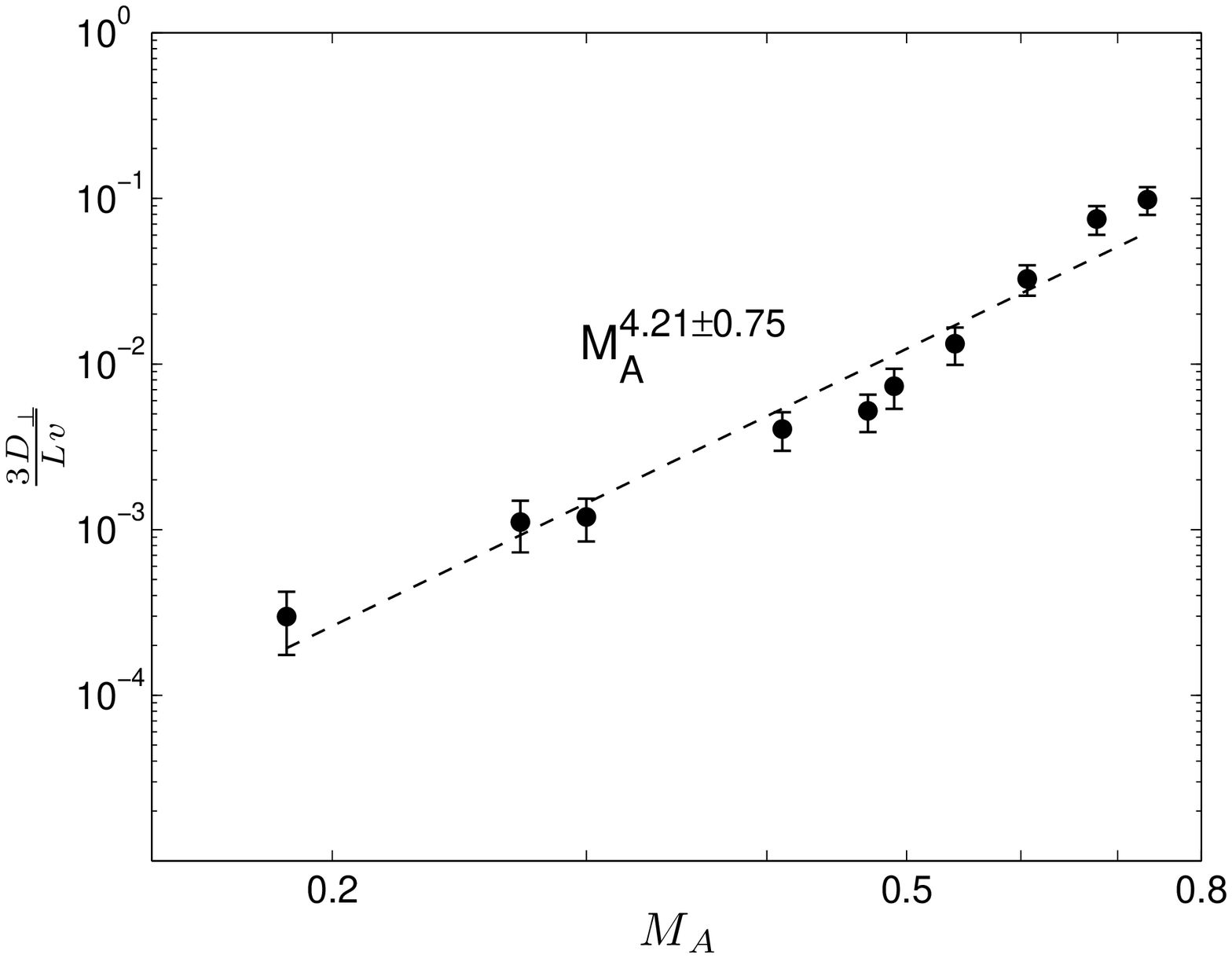}}
   \subfigure[]{
   \includegraphics[width=0.48\textwidth]{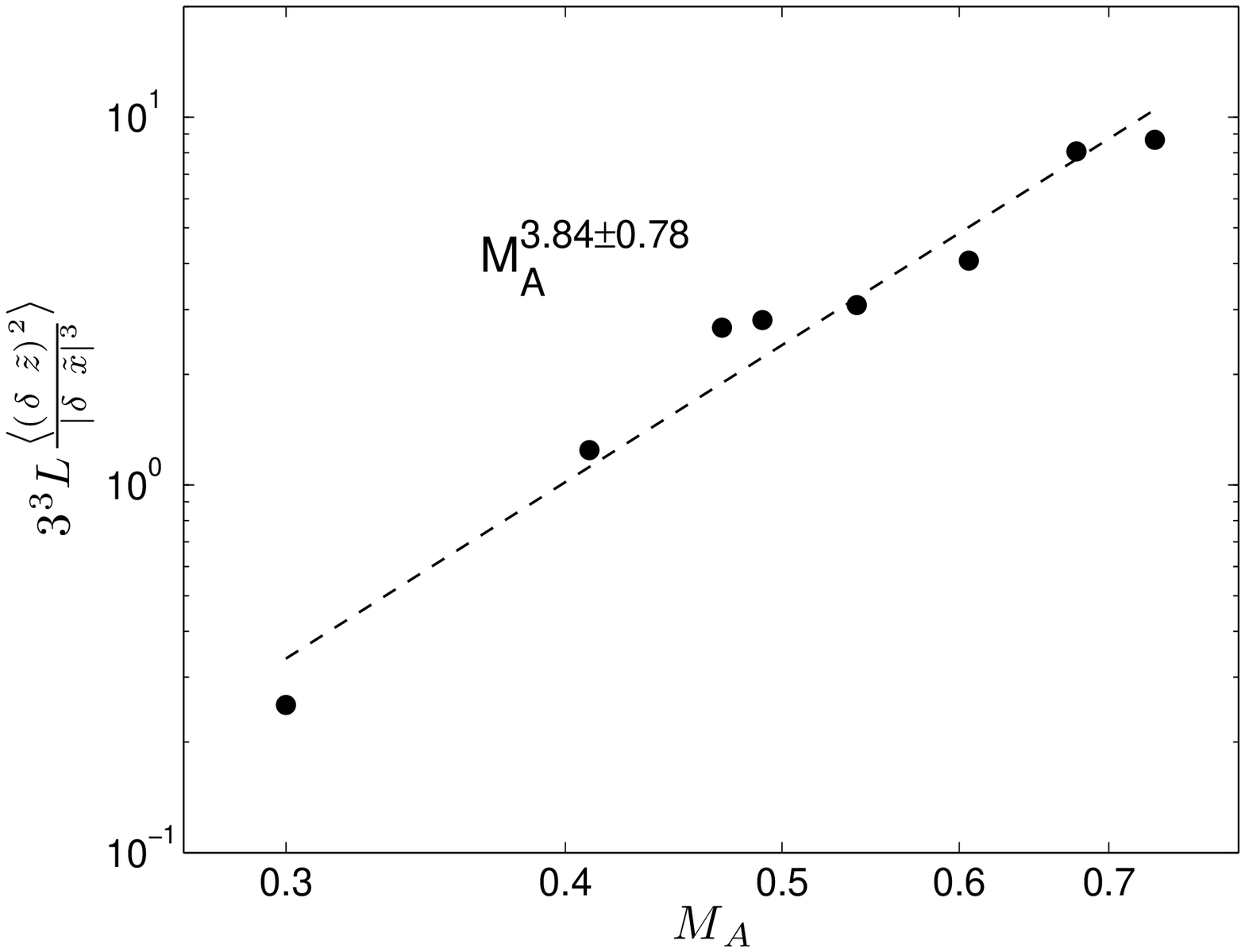}}
   \caption{\small (a) $\frac{3D_\perp}{Lv}$ as a function of $M_A$.The dashed line shows the best fit to the numerical results (filled circles); (b)  the normalized separation of particle trajectories as a function of $M_A$. The dashed lines are the best fits to the data. \citep*[from][]{Xu_Yan}.}
   \label{fig6}
\end{figure}

As shown in Fig.\ref{fig6}, the test particle simulations conducted by \cite{Xu_Yan} confirm the dependence of $M_A^4$ instead of the $M_A^2$ scaling in, e.g., 
\citet{Jokipii1966}. This is exactly due to the anisotropy of the Alfv$\acute{\rm e}$nic turbulence. In the case of sub-Alfv$\acute{\rm e}$nic turbulence, the eddies become elongated along the magnetic field from the injection scale of the turbulence 
(\citealt{Lazarian06}; YL08). 
The result indicates that CR perpendicular diffusion depends strongly on $M_A$ of the turbulence, especially in magnetically dominated environments, e.g., the solar corona.

\section{Superdiffusion on small scales}

\begin{figure}[htbp] 
    \centering
   \subfigure[]{
   \includegraphics[width=0.45\textwidth]{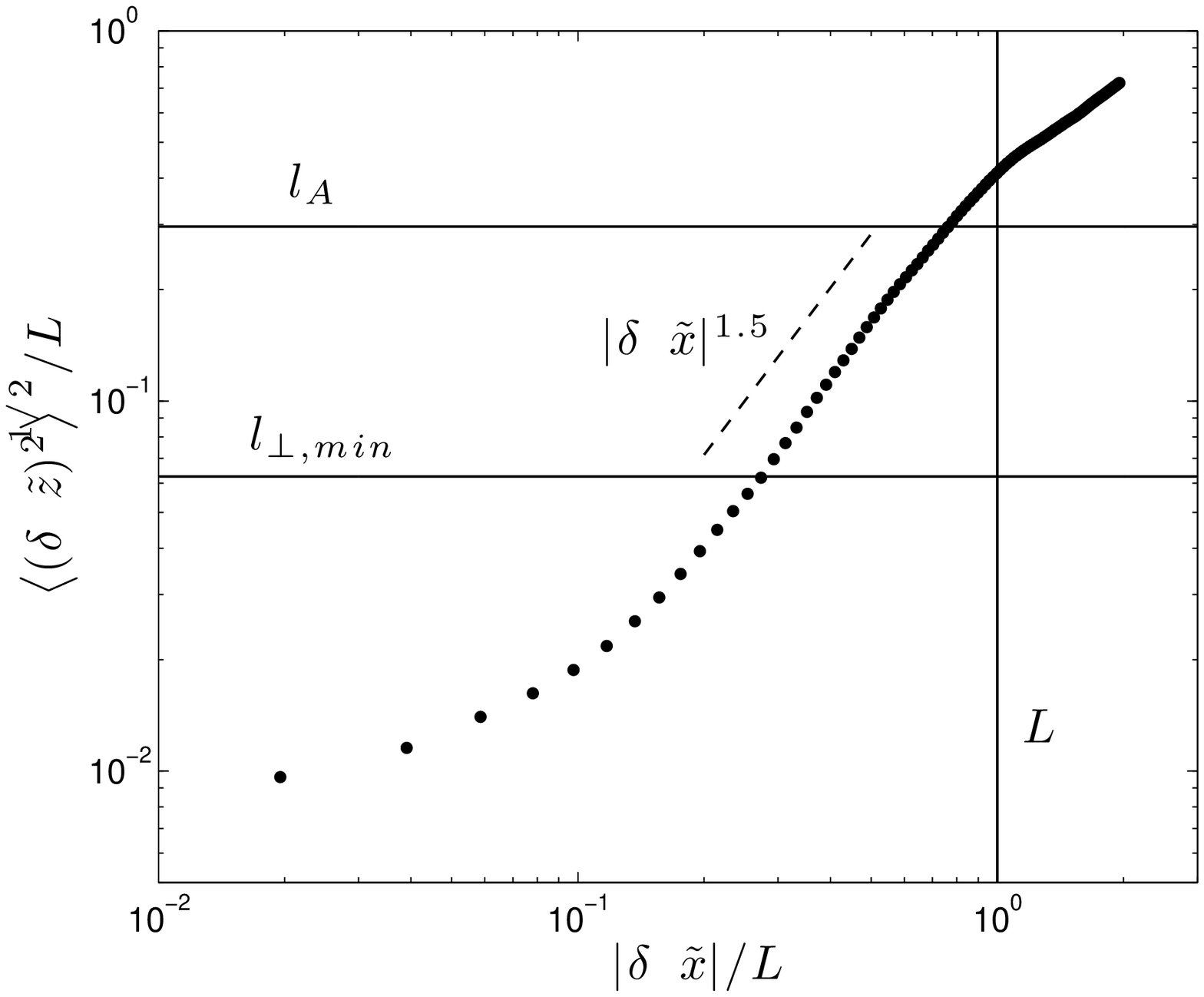}}
    \centering
   \subfigure[]{
   \includegraphics[width=0.45\textwidth]{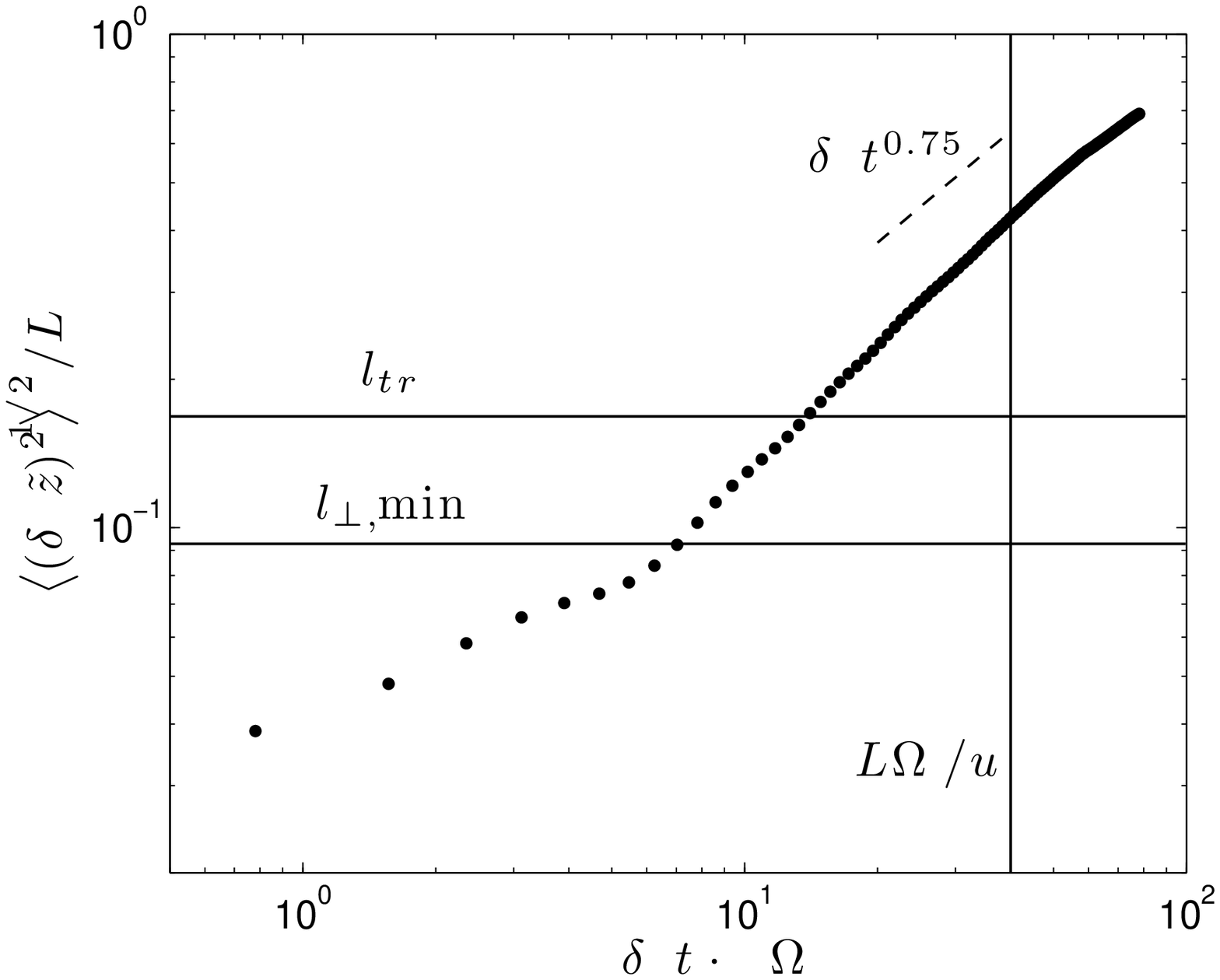}}
   \caption{\small The normalized separation of particle trajectories $\left \langle (\delta \tilde{z})^2 \right \rangle^{1/2}/L$ (a) vs. $|\delta \tilde{x}|/L$;
   (b)  vs. $\Omega \cdot \delta t$ for $\lambda_\parallel < L$. The vertical line denotes the time for particles to travel $L$ along the direction of magnetic field\citep*[from][]{Xu_Yan}.}
   \label{figev}
\end{figure}

The diffusion of CR on the scales $\ll L$ is different and it is determined by how fast field lines are diverging away from each other. The mean separation of field lines $\delta x$ is proportional to the 3/2 power of distance $[\delta z]^{3/2}$ \citep{LV99, Lazarian06}, same as Richardson diffusion in the case of hydrodynamic turbulence \cite[see][]{ELV11}. Following the argument, we showed in \cite{YL08} that the cosmic ray perpendicular transport is superdiffusive. The reason is that there is no random walk on small scales up to the injection scale of strong MHD turbulence ($LM_A^2$ for $M_A < 1$
 and $l_A$ for $M_A>1$). In the case of sufficiently strong scattering, particles' mean free path becomes smaller than the scale of interest so that they diffuse along the field lines which are diverging away $[\delta z]^{3/2}$ and the separation of particle trajectories therefore goes as $\propto t^{3/4}$ (see Fig.\ref{figev}b).
 
\subsection{Superdiffusion and shock acceleration}

The super diffusion on small scales has important consequences for shock acceleration. For shock acceleration, the crucial element is the confinement of particles by magnetic perturbations at upstream and downstream regions. The first order Fermi process can be only realized if particles can be scattered back and forth consecutively before they escape the shock front. Ultimately both the acceleration rate and the maximum energy attainable are determined by the diffusion properties of particles in the local turbulence. 

There have been arguments that the perpendicular shock is much more efficient accelerator than the parallel one based on the suppressed diffusion   perpendicular to the magnetic field. And perpendicular shock thereby has been invoked to remedy the problem of insufficient acceleration by parallel shocks at places like termination shock \citep{Jokipii87}. In the presence of turbulence and superdiffusion, nevertheless, the difference between the parallel shock and perpendicular shock becomes diminished. Only when local turbulence is generated at the shock with sufficiently small integral scale, the super diffusion which only happens on scales smaller than the injection scale can be neglected and the shock acceleration can be more effective. We refer the readers to \cite{LY13} for the detailed discussions.
 
\section{Summary}
In the paper above, we presented current understanding of cosmic ray transport in tested model of turbulence. Nonlinear approach was employed along with the numerical testings. We showed that 
\begin{itemize}
\item Compressible fast modes are most important for CR scattering. CR transport therefore varies from place to place.
\item Mirror interaction is essential for pitch angle scattering (including 90 degree).
\item Subdiffusion does not happen in 3D turbulence.
\item On large scales, CR perpendicular diffusion is suppressed by
$M_A^4$ 
\item On small scales, CR transport is super-diffusive, has a dependence
of $M_A^4$ in sub-Alfvenic turbulence.
\item Implications are wide, from shock acceleration to turbulent reconnection. 
\end{itemize}

\bibliography{yan}


\end{document}